# Mesophase behaviour of binary mixtures of bent-core and calamitic compounds

M. Cvetinov[a], D. Obadović[a], M. Stojanović[a], D. Lazar[a], A. Vajda[b], N. Éber[b], K. Fodor-Csorba[b]
and I. Ristić[c]

[a] Department of Physics, Faculty of Sciences, University of Novi Sad, Novi Sad, Serbia
[b] Institute for Solid State Physics and Optics, Wigner Research Centre for Physics, Hungarian Academy of Sciences, Budapest, Hungary
[c] Technological Faculty, University of Novi Sad, Novi Sad, Serbia

The mesophase behaviour of binary mixtures of bent-core and calamitic liquid crystals is presented. The nematogenic 4,6-dichloro-1,3-phenylene bis[4'-(10-undecen-1-yloxy)-1,1'-biphenyl-4-carboxylate] (**I**) was the banana-shaped component. As the calamitic compound ethyl 4'-(9-decen-1-yloxy)-1,1'-biphenyl-4-carboxylate (**II**), similar to one arm of the bent-core molecule, was used which exhibits smectic phases in a wide temperature range. A total of 6 mixtures with different compositions were prepared and studied by polarising optical microscopy, differential scanning calorimetry and X-ray diffraction on non-oriented samples. In the mixtures a nematic phase is not concomitant with smectic A phase, and the temperature range of both phases highly depends on the concentration of the comprising compounds. Lowered melting temperatures have been observed for all mixtures with respect to that of the pure compounds. Unforeseen finding is the induction of a monotropic SmC phase in mixtures with lowest concentration of the bent-core compound. Semi-empirical quantum-chemical calculations have also been performed. Based on the calculated molecular conformation, as well as on collected X-ray diffraction data, a model for a possible self-assembly of the banana-shaped and calamitic compounds is proposed.

**1. Introduction**

Bent-core (banana-shaped) compounds represent a new class of thermotropic liquid crystals with a nonconventional architecture and an ability to exhibit mesomorphic properties (banana phases B1–B8) different from those of classical liquid crystals [1-5]. Inspired by unusual properties of their mesophases, bent-core compounds have been investigated intensively in the last decade [6-9].

Dimers of calamitic mesogenic units are also able to form bent structures if their flexible spacer has an odd number of carbon units. Such dimers (especially if they are asymmetric) show a tendency to form intercalated smectic phases with tilt directions alternating in neighbouring smectic layers [10,11]. Recently an intercalated $B_6$ banana phase [1,12] formed by symmetric calamitic dimers has also been reported [13,14].

In comparison to specific banana phases, nematic phases are relatively rare amongst mesophases of bent core compounds [15]. The occurrence of nematic phases requires either molecules with extended aromatic cores and relatively short terminal chains or a reduction of the molecular bent. One way to achieve this is to introduce a large substituent at the 4-position of the central phenylene ring, thus widening the bend angle of the unsubstituted 1,3-phenylene ring and rendering the molecules less prone to polar packing in layers [16]. As the shape of the bent-core molecules is strongly biaxial, bent-core compounds are candidates for forming a bulk biaxial nematic phase. So far, however, these expectations have not been met; despite of several claims the occurrence of a biaxial nematic phase in a thermotropic liquid crystal could not been proved unambiguously [17].

The uniaxial nematic phase is particularly important among liquid crystalline phases, due to its application in liquid crystal displays (LCDs). Lowering the transition temperatures of the nematic phase and extension of its temperature range has always been an important aim of studies, as these liquid crystalline materials often have high clearing points. In order to respond to these demands, the required liquid crystalline properties can be reached rather by mixing compounds with various molecular shapes and properties than by looking for a single compound with all required properties. Mixing compounds of different molecular structures has proven to be a useful tool to achieve lower transition temperatures in binary systems. Although bent-core compounds exhibit limited miscibility among themselves, a way to eschew this problem is to mix them with calamitic compounds [18-24], which could lead to unusual self-assemblies [25,26].

Recently synthesis of banana–calamitic dimers or trimers opened an alternative way of tuning the physical properties [27,28]. These oligomers may in some sense be regarded as analogues of mixtures with specific concentrations: dimers correspond to a 50–50 mol%, while trimers correspond to a 66.6–33.3 mol% or 33.3–66.6 mol% binary mixture of a bent-core and a calamitic compound. One has, however, to keep in mind that in the oligomers the different mesogenic units are connected with a spacer group of fixed length (whatever flexible it is), while in the mixtures the individual molecules are free to diffuse. Therefore the self-assembly provided by the oligomers should not necessarily be identical with that of a mixture with matching concentration.

In the present article we report on miscibility studies on the binary system of the bent-core compound 4,6-dichloro-1,3-phenylene bis[4'-(10-undecen-1-yloxy)-1,1'- biphenyl-4-carboxylate] (**I**) [16] and the rod-like compound ethyl 4'-(9-decen-1-yloxy)-1,1'-biphenyl-4-carboxylate (**II**) [29] whose syntheses were described earlier. Chemical structures of the studied compounds are shown in Figure 1. As seen in Figure 1, **II** is almost identical with one arm of the bent core compound.

Our aim was to lower the phase transition temperatures and to study the properties of such binary systems. We report on polarising optical microscopy (POM) and differential scanning calorimetry (DSC) studies as well as on X-ray measurements of the mixtures.

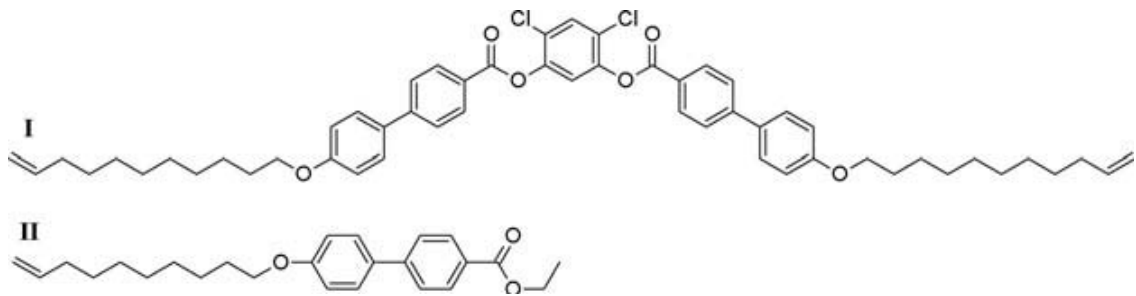

Figure 1. Chemical structures of the bent-core **I** and the calamitic **II** compounds.

## 2. Experimental

Sequences of phases and phase-transition temperatures were determined from the characteristic textures and their changes observed in a polarizing optical microscope Carl Zeiss Jena equipped with a hot-stage for the controlled heating and cooling of the sample. For differential scanning calorimetry (DSC) analysis the Du Pont Instrumental Thermal Analyzer 1090 910 was used. During the measurements a heating/cooling rate of 5°C/min was applied. Combination of microscopy with DSC studies enabled us to identify mesophases and to construct the phase diagrams.

In order to obtain more structural information, non-oriented samples were investigated by X-ray diffraction in a transmission geometry using a conventional powder diffractometer, Seifert V-14, equipped with an automated high-temperature kit Paar HTK-10. Diffraction was detected at the $CuK_\alpha$ radiation of 0.154059 nm.

Table 1. Phase transition temperatures [°C] and enthalpies [J/g] (in square brackets) for the pure compounds and the binary mixtures obtained by DSC in heating (• the phase exists).

| Code | Mol% of I | Cr1 | T(°C) | Cr2 | T(°C) | SmE | T(°C) | SmC | T(°C) | SmA | T(°C) | N | T(°C) | I |
|---|---|---|---|---|---|---|---|---|---|---|---|---|---|---|
| **II** | 0 | • | 76.4 [4.1] | | | • | 86.4 [3.7] | | | • | 100.4 [18.0] | | | • |
| **Mix1** | 8.7 | • | 56.0 [1.73] | • | 67 [0.94] | • | 78.5 [0.13] | | | • | 96.5 [4.2] | | | • |
| **Mix2** | 13.3 | • | 56.4 [5.86] | • | 66 [0.81] | • | 79 [0.41] | | | • | 89.7 [1.88] | | | • |
| **Mix3** | 21.1 | • | 46 [0.78] | • | 56.8 [17.8] | • | 65 [0.49] | | | | | • | 79.8 [1.19] | • |
| **Mix4** | 30.4 | • | 46.5 [0.6] | • | 56.2 [16.9] | | | | | | | • | 80.4 [1.22] | • |
| **Mix5** | 42.6 | • | 46.6 [0.67] | • | 56.2 [21.1] | | | | | | | • | 84.1 [1.1] | • |
| **Mix6** | 63.6 | • | 57 [53.4]* | • | 68 | | | | | | | • | 92.7 [1.99] | • |
| **I** | 100 | • | 75.8 [58.7] | | | | | | | | | • | 104.9 [1.4] | • |

Note: *The close transitions are overlapping; only the overall enthalpy could be given.

Table 2. Phase transition temperatures [°C] and enthalpies [J/g] (in square brackets) for the pure compounds and the binary mixtures obtained by DSC in cooling (• the phase exists).

| Code | Mol% of I | Cr1 | T(°C) | SmE | T(°C) | SmC | T(°C) | SmA | T(°C) | N | T(°C) | I |
|---|---|---|---|---|---|---|---|---|---|---|---|---|
| **II** | 0 | • | 73.6 [3.29] | • | | | 84.1 [2.91] | • | | | 96.4 [19.0] | • |
| **Mix1** | 8.7 | • | 59.6 [0.7] | • | 63 [0.23] | • | 78.5 [0.35] | • | | | 87.3 [4.26] | • |
| **Mix2** | 13.3 | • | 59.4 [1.46] | • | 63.8 [-]** | • | 78.1 [0.70] | • | | | 81.3 [-]** | • |
| **Mix3** | 21.1 | • | 58.4 [-]** | • | 64.7 [0.51] | • | | | 73.5 [-]** | • | 78.5 [1.15] | • |
| **Mix4** | 30.4 | • | | | | | | | 55.4 [-]** | • | 79.6 [1.4] | • |
| **Mix5** | 42.6 | • | | | | | | | 53.7 [-]** | • | 83.1 [1.05] | • |
| **Mix6** | 63.6 | • | | | | | | | 44.2 [54.1] | • | 91.4 [2.05] | • |
| **I** | 100 | • | | | | | | | 60.9 [54.7] | • | 102.7 [1.84] | • |

Note: ** The intensity of the DSC peak is comparable with the sensitivity; reliable enthalpy data are not available

## 3. Results and discussion

The goal of the present work was to test the miscibility of the bent-core compound **I** with the rod-like material **II**, and to investigate the mesomorphic behaviour of their binary mixtures. For the detailed study, six mixtures, **Mix1** to **Mix6**, have been prepared with 18, 26, 38, 50, 63 and 80 wt% (8.7, 13.3, 21.1, 30.4, 42.6 and 63.6 mol%) of the bent-core compound **I**, respectively (see also Tables I and II). Mixtures under investigation were stable, showing no signs of segregation after one month storage period.

The pure compounds as well as their mixtures were investigated by DSC technique. Figure 2 shows representative DSC curves for all mixtures obtained in a first cooling and second heating cycle.

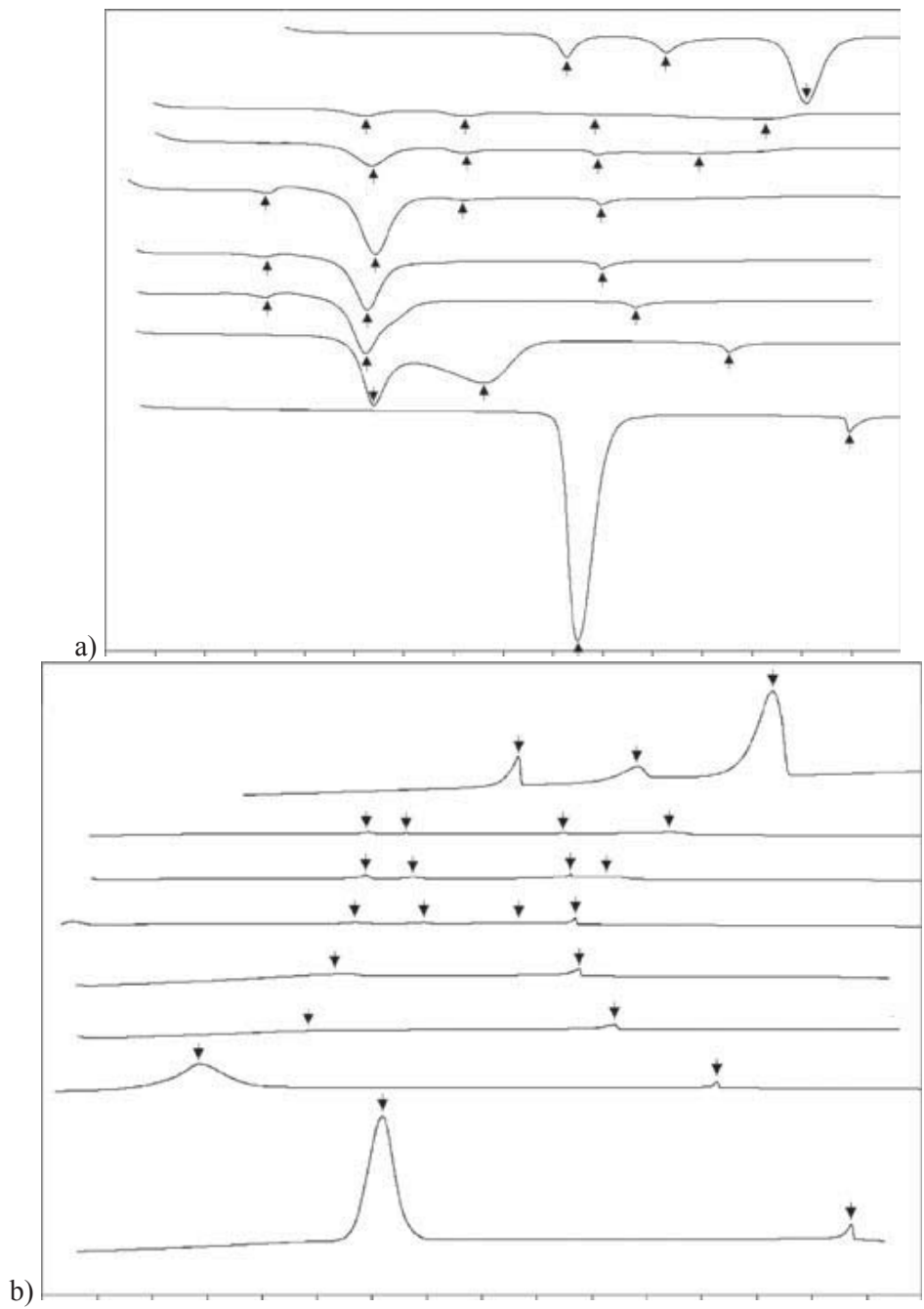

Figure 2. DSC plots of the pure compounds and the binary mixtures in (a) heating, (b) cooling.

The phase transition temperatures and transition enthalpies of the studied compounds evaluated from the DSC curves are summarized in Tables I and II for heating and cooling, respectively. Note that the N–I transition enthalpy obtained in cooling (1.58 kJ/mol) of the bent-core compound **I** is at the lower end of enthalpy value range of typical calamitic nematics [13,30]. This might be attributed to the bent molecular shape; there are indications that increasing the molecular biaxiality reduces the entropy and enthalpy changes at the N–I phase transition [31].

While the bent-core compound **I** was purely nematogenic, the calamitic compound **II** exhibited SmA and SmE but no nematic phases. Mixtures with high concentration of the calamitic component (**Mix1** and **Mix2**) showed the highest morphological richness, as they preserved the SmA and SmE phases of compound **II** and in addition they possessed an induced SmC phase too. The widest temperature range for the SmA phase (12.3°C in cooling) was observed in the pure calamitic compound; the widest SmC range (15.5°C in cooling) was found in **Mix1**. At increasing the content of the bent-core component, first the SmA phase was replaced by the nematic one (**Mix3**); then in mixtures with equal or higher concentration of the banana compound **I** only nematic mesophase was found. The temperature range of the nematic phase was widest (47.2°C) in **Mix6**. For a better illustration of the polymorphism the binary phase diagram of the system is also provided in Figure 3(a) and 3(b).

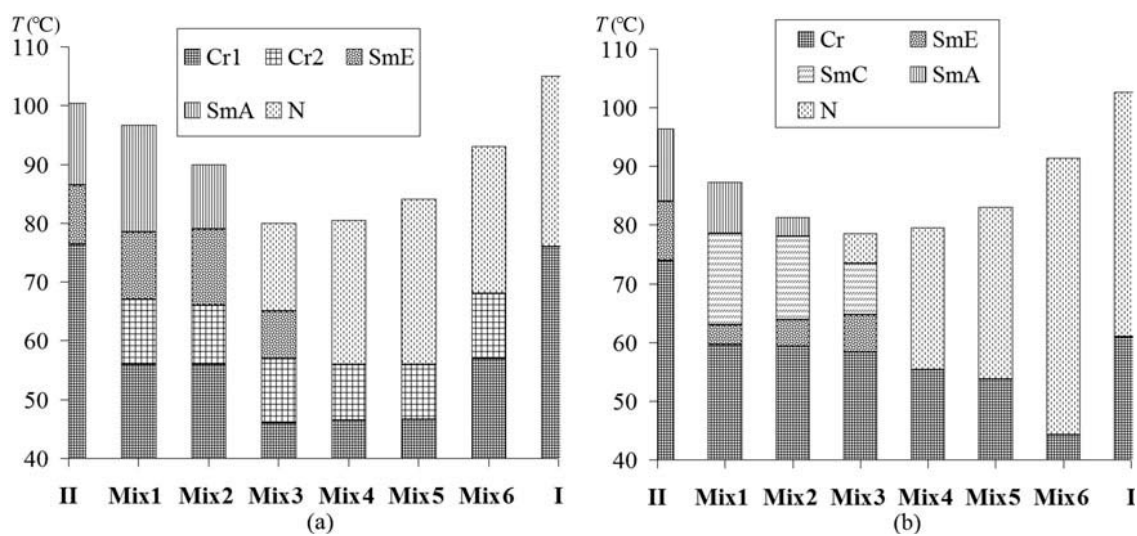

Figure 3. Bargraphs indicating the phase sequences of the binary system composed of compounds **I** and **II** obtained by DSC and/or polarising optical microscopy (POM): (a) in heating; and (b) in cooling.

The identification of the phases (indicated in Figure 2, Figure 3 and in Tables 1 and 2) was made by POM studies. Microphotographs of some characteristic textures of the various mesophases obtained in cooling of non-homogenously aligned planar samples are presented in Figure 4.

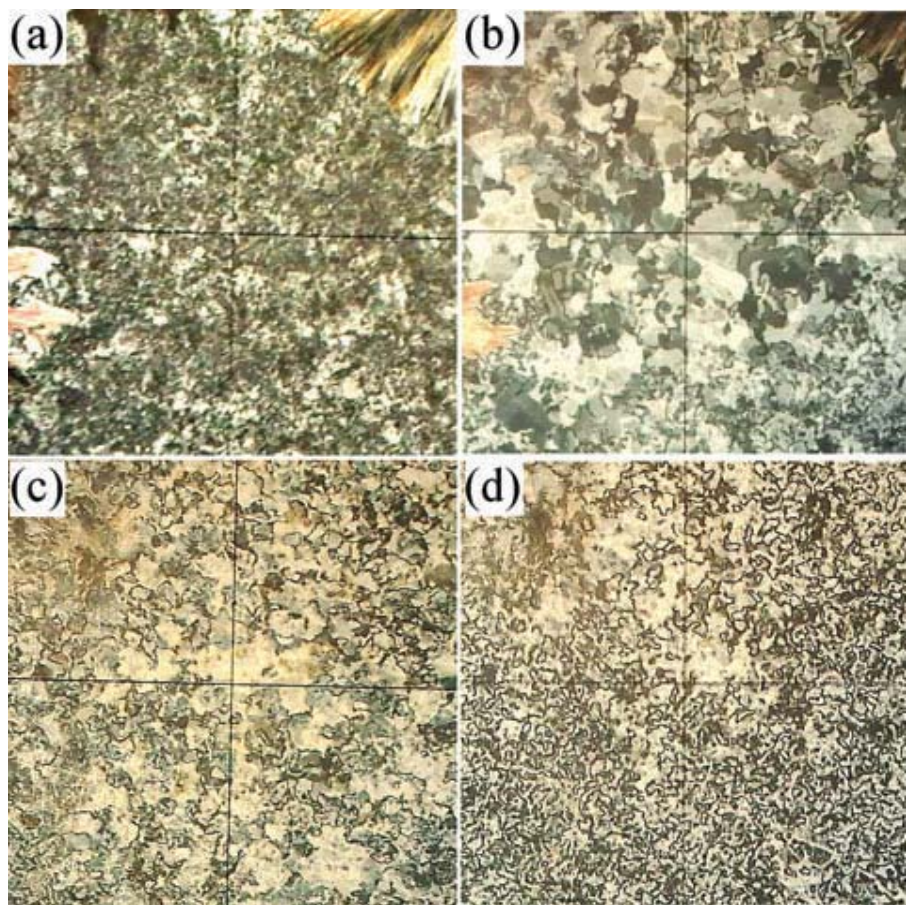

Figure 4. Micrographs of unoriented samples of **Mix3** in cooling in: (a) crystalline phase at 25°C; (b) smectic SmE phase at 60°C, mosaic texture; (c) smectic SmC phase at 66°C, blurred Schlieren texture; (d) nematic phase at 75°C, Schlieren texture. All micrographs are 350 μm wide.

Diffraction studies were carried out on the pure compounds **I** and **II**, as well as on the binary mixtures. Two parameters characteristic for molecular packing, the thickness of the smectic layers d (if layers exist) and the average intermolecular distance D between the long axes of neighbouring molecules [32-35], could be determined from the positions of the small angle and wide-angle diffraction peaks, respectively. The evaluation was based on the Bragg law: $n\lambda = 2d \sin\Theta$, where $\lambda$ is the radiation wavelength, $\Theta$ is the scattering angle and d is the repetition distance to be determined. These results are summarised in Table III. Second order reflection peaks (n=2) are not included in the table. Calibration of the X-ray set-up was performed using a platinum sample, measuring its two most intense diffraction peaks with well-established $2\theta$ values [36]. The resulting corrections for zero shift and sample displacement errors were included in the table.

Table 3. Molecular parameters of the investigated mixtures for all observed phases at a fixed temperature T (°C): angles corresponding to the reflection peaks 2Θ (degrees), effective layer thickness d (in nm; error of measurements was $\sigma_d \approx 0.001$ nm), average repeat distance D (in nm; error of measurements was $\sigma_D \approx 0.002$ nm).

| Mixture | T (°C) | 2θ (°) | d (nm) | D (nm) |
|---|---|---|---|---|
| **II** | 110(I) | 19.4 | | 0.457 |
| | 90(SmA) | 3.3 | 2.675 | |
| | | 20.55 | | 0.432 |
| | 75(SmE) | 3.3 | 2.675 | |
| | | 20.25 | | 0.438 |
| | | 22.7 | | 0.391 |
| **Mix1** | 97(I) | 18.5 | | 0.479 |
| | 79(SmA) | 3.3 | 2.675 | |
| | | 19.0 | | 0.467 |
| | 66(SmC) | 3.7 | 2.386 | |
| | | 19.0 | | 0.467 |
| | 62(SmE) | 3.3 | 2.675 | |
| | | 19.2 | | 0.462 |
| | | 20.8 | | 0.427 |
| | | 22.5 | | 0.395 |
| **Mix2** | 110 (I) | 18.6 | | 0.476 |
| | 79 (SmA) | 3.3 | 2.675 | |
| | | 18.8 | | 0.471 |
| | 70(SmC) | 3.7 | 2.386 | |
| | | 18.9 | | 0.469 |
| | 63(SmE) | 3.25 | 2.715 | |
| | | 22.1 | | 0.402 |
| **Mix3** | 100(I) | 18.0 | | 0.492 |
| | 78(N) | 18.5 | | 0.479 |
| | 68(SmC) | 3.75 | 2.354 | |
| | | 18.9 | | 0.469 |
| | 63(SmE) | 3.3 | 2.675 | |
| | | 19.7 | | 0.450 |
| | | 21.9 | | 0.405 |
| **Mix4** | 110(I) | 19.0 | | 0.467 |
| | 77(N) | 19.3 | | 0.459 |
| **Mix5** | 100(I) | 19.0 | | 0.467 |
| | 75(N) | 19.5 | | 0.455 |
| **Mix6** | 103(I) | 19.2 | | 0.462 |
| | 79(N) | 19.7 | | 0.450 |
| **I** | 110(I) | 19.3 | | 0.459 |
| | 90(N) | 20.0 | | 0.444 |

In Figure 5(a) and 5(b) we present typical diffraction spectra for each phase for the pure calamitic compound **II** as well as for the mixture **Mix1**, in order to demonstrate the change occurring in the phase transitions. In both diffractograms, the SmE phase is primarily characterised by increased scattering at large angles in proximity of 2θ=20.5° which is due to the hexatic ordering. In the SmA phase of both the pure compound **II** and **Mix1**, first order reflection peak appears at the small angle 2θ=3.3° and second order reflection peak appears at angle 2θ=6.6°, corresponding to the thickness of smectic layers of d=2.67nm, which is in agreement with the calculated length of compound **II**. An induced SmC phase was found in **Mix1,** which is characterized by the peak at the small angle of 2θ=3.7°. The most probable explanation for this reflection peak is that it indicates the thickness of the smectic layers of d=2.39nm. The isotropic phase is characterised by a broad diffusion peak which appears in the range of 2θ=12–26° corresponding to the lateral distance between the long molecular axes.

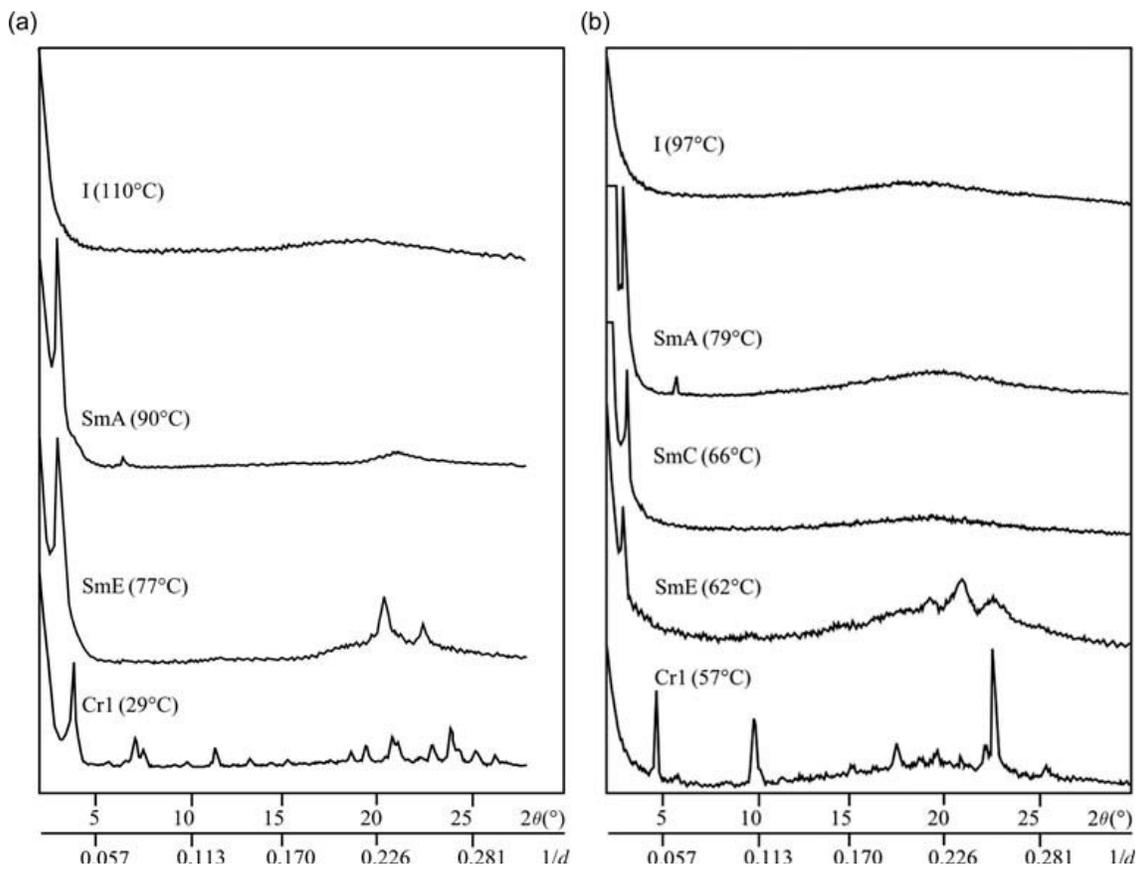

Figure 5. X ray diffractogram of (a) pure **II** compound and (b) **Mix1** in cooling.

Molecular models were constructed in order to give us an insight into the problematic self-assembly of the banana-shaped and calamitic materials in the mesophase. Computation was performed with using an RM1 parameterization of the semi-empirical method [37]. The results of computation were then compared with the X-ray measurements.

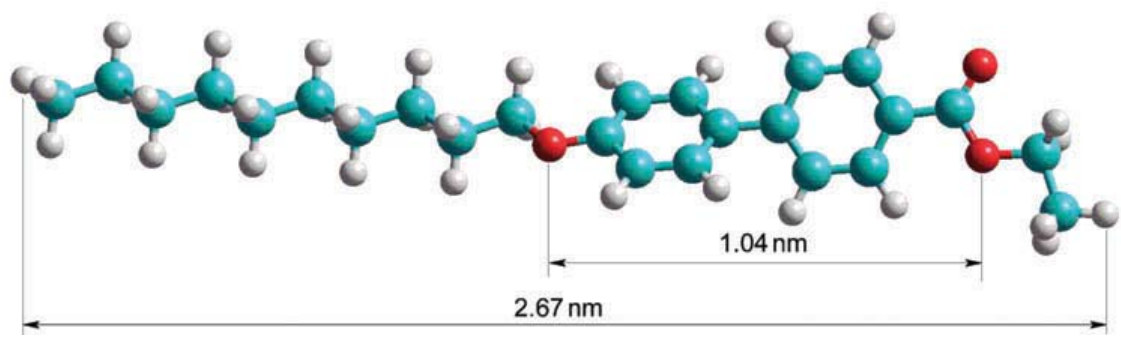

Figure 6. Optimized geometry of compound **II.**

The biphenyl ring moiety has the smallest torsion angles (37-45°) when the rings are unsubstituted or substituted with small atoms such as fluorine or oxygen [38]. The torsion angles increase with increasing bulkiness of the substituents. In case of compound **II,** geometry optimization using the Polak-Ribiere algorithm yielded a torsion angle of 49.44° between phenyl rings, a total molecular length of 2.67nm and length of the rigid core of the molecule of 1.04nm, as shown in Figure 6. Nevertheless, the calamitic molecule is not completely linear but has a slightly bent structure which might be helpful for the self-assembly with the bent-core compound in the mesophase.

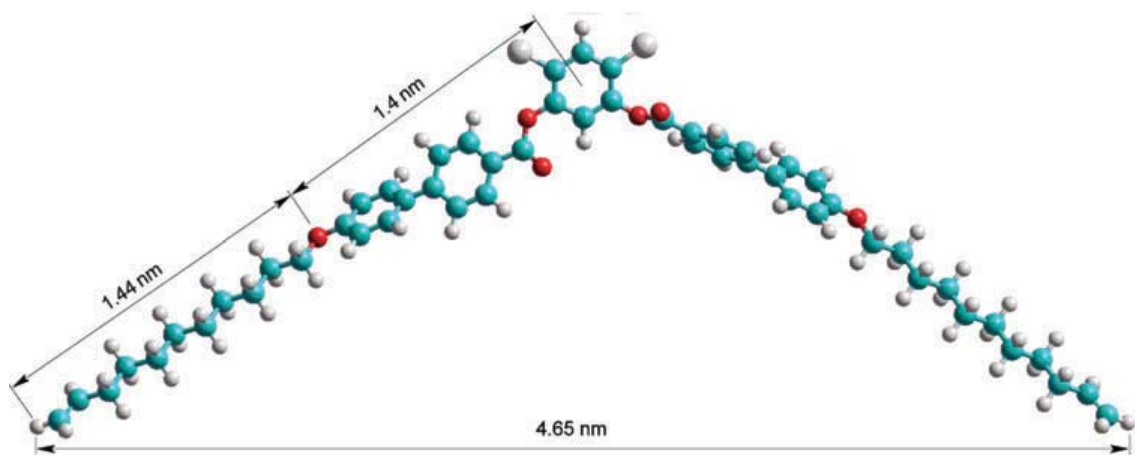

Figure 7. Optimized geometry of compound **I.**

Minimum energy conformation of compound **I** is shown in Figure 7. The density of electrons is highest in the vicinity of atoms characterized by high electronegativity (principally at oxygen atoms), followed by slightly less density in the vicinity of chlorine atoms. The calculated bending angle between the two arms of the molecule is 121.32°, slightly less than the previously reported value [16]. The angle between the two neighbouring phenyl rings in both arms of the molecule is 44.1°. Recalculation using RM1 parameterization yielded 4.65nm for the molecular length of compound **I**; the linear length of the half of its rigid molecular core is 1.4nm and the linear length of each alkyl chain is 1.44nm.

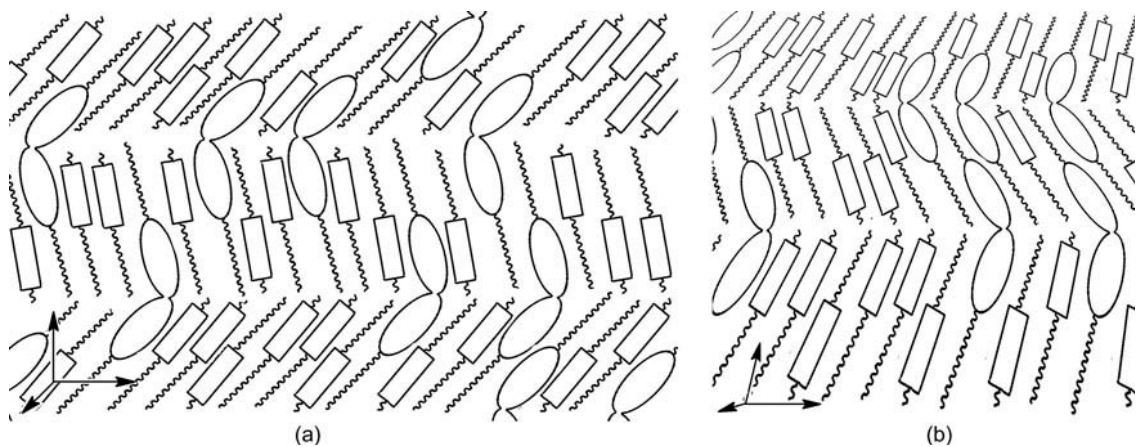

Figure 8. Proposed molecular packing of molecules in smectic layers of the SmC phase of **Mix3,** where a) bent-core molecules are tilted by fixed amount in their molecular plane which remains perpendicular to the layers; b) bent-core molecules are leant by fixed amount perpendicular to their molecular plane.

Knowing the molecular sizes one can attempt to construct a model for the self-assembly of molecules in the experimentally found SmC phase of **Mix3**. Due to the large difference in the molecular sizes and thus in the molar masses of **I** and **II**, the composition of **Mix3** (38 wt% of **I**) actually corresponds to a molar concentration of 21.1 mol% of compound **I**, i.e. there are 3.7 times more calamitic molecules than bent-core ones in the mixture. The X-ray data in Table 3 clearly show that the smectic layer thickness is shorter than the length of compound **II**. Hence the bent-core molecules are expected to intercalate the smectic layers, causing a tilting of the calamitic molecules. If the bent-core molecules were normal to the smectic layers, this structure (similarly to an intercalated SmA or a $B_6$ phase) would yield an orthogonal (SmA-like) phase. In the SmC phase thus the molecules of **I** should be tilted with respect to the layer normal. In Figure 8 we suggest two possible configurations of molecules for such a tilted smectic C phase of **Mix3**. In Figure 8(a) molecules are tilted in the molecular plane of the banana compound, while in Figure 8(b) the molecules lean in a perpendicular direction. In both alternatives as well as in their possible combination the calamitic (**II**) and the bent-core molecules (**I**) retain their nonpolar order in the smectic layers, due to the equal probability of molecules pointing in any of two possible directions (up and down, or left and right). The two suggested models are not distinguishable by our experimental techniques, neither by polarizing microscopy nor by X-ray powder diffraction; we leave the question of correct molecular packing open for future experimental studies.

## 4. Conclusions

The present studies were performed with the aim of contributing to the understanding of how mixing of bent-core and rod-like molecules affect the mesomorphic properties. Based on POM, DSC and X-ray measurements on several mixtures, we have found that the polymorphism of the pure calamitic component **II** is fully preserved only in mixtures with the highest concentrations of compound **II** (**Mix1** and **Mix2**). Interesting and unforeseen finding is the induction of a monotropic SmC phase that is observed in **Mix1, Mix2** and **Mix3.** The nematic phase remains detectable for mixtures with equal or higher concentration of banana compound **I**. Therefore, complex mesophase behaviour existed over a broad compositional range in the mixtures and could be extended close to room temperature. The results suggest that combining conventional calamitics with bent-core mesogens with an appropriate molecular design may be a tool to tune the phase behaviour and properties of different liquid crystal mixtures.

**Acknowledgements**

This work was partly supported by the research Grant No. OI171015 from the Ministry of Education and Science of the Republic of Serbia, by the Hungarian Research Fund OTKA K81250 and the SASA-HAS bilateral scientific exchange project #9.